# Link between SARS-CoV-2 emissions and airborne concentrations: closing the gap in understanding

G. Buonanno[1,2], A. Robotto[3], E. Brizio[3], L. Morawska[2], A. Civra[4], F. Corino[3], D. Lembo[4], G. Ficco[1], L. Stabile[1,*]

[1] Department of Civil and Mechanical Engineering, University of Cassino and Southern Lazio, Cassino, FR, Italy
[2] International Laboratory for Air Quality and Health, Queensland University of Technology, Brisbane, Qld, Australia
[3] Environmental Protection Agency of Piedmont (ARPA Piemonte), Italy
[4] Dept. of Clinical and Biological Science, Azienda Ospedaliero-Universitaria San Luigi Gonzaga, University of Turin, Italy

**Abstract**

The question of how SARS-CoV-2 is transmitted remains surprisingly controversial today, especially with reference to airborne transmission. In fact, despite a large body of scientific evidence, health and regulatory authorities still require direct proof of this mode of transmission. To close this gap, we measured the saliva viral load of SARS-CoV-2 of an infected subject located in a hospital room, as well as the airborne SARS-CoV-2 concentration in the room resulting from the person breathing and speaking. As the next step, we simulated the same scenarios to estimate the concentration of RNA copies in the air through a novel predictive theoretical approach. Finally, we conducted a comparative analysis (i.e. a metrological compatibility analysis) of the differences between the experimental and theoretical results by estimating the uncertainties of these two approaches.

Our results showed that for an infected subject's saliva load ranging between $2.4 \times 10^6$ and $5.5 \times 10^6$ RNA copies mL$^{-1}$, the corresponding airborne SARS-CoV-2 concentration was not detectable when the person was breathing, but was 16.1 (with an uncertainty of ± 32.8) RNA copies m$^{-3}$ when speaking. The application of the novel predictive estimation approach provided average concentrations of 3.2 (uncertainty range of 0.2–8.3) and 18.5 (uncertainty range of 4.5–43.0) RNA copies m$^{-3}$ for breathing and speaking scenarios, respectively, thus confirming that for the breathing scenario, the airborne RNA concentration would be undetectable, being below the minimum detection threshold of the experimental apparatus (< 2 RNA copies m$^{-3}$).

To summarize, the experimental analysis performed here provided significant evidence to close the gap in understanding airborne transmission, given that the airborne SARS-CoV-2 concentration was shown to be directly related to the SARS-CoV-2 emitted. Moreover, the theoretical analysis presented here (and validated through the metrological compatibility analysis) was shown to quantitatively link the airborne concentration to the emission.

## 1       Introduction

The transmission mode of SARS-CoV-2 has divided experts and public health communities since the beginning of the COVID-19 pandemic. The main object of contention was the relevance of the airborne transmission route by inhalation of virus-laden respiratory particles emitted by an infected subject. In this context, the adoption of different terminology by scientific and medical communities, assigning different meanings to the terms "aerosol" and "droplets" (Randall et al., 2021), did not help. To avoid misinterpretation in the paper, we use the term "particles" rather than "aerosols" or "droplets". While interdisciplinary expert groups warned of the need to recognize the importance of the airborne transmission route (Greenhalgh et al., 2021; Morawska et al., 2020; Morawska and Cao, 2020), public health authorities were reluctant to accept this. In fact, after more than a year into the pandemic, the World Health Organization (WHO), the US Centers for Disease Control and Prevention (CDC), and the Italian National Institute of Health (ISS) had only partially recognized the importance of airborne transmission (WHO, 30 April 2021; US CDC 7 May 2021, ISS 25 May 2021). Indeed, public health authorities argued that to accept the relevance of airborne transmission, direct experimental evidence of its occurrence was needed. Interestingly, such verification has never been required for other transmission routes, such as droplet (larger particles) or fomite transmission (Greenhalgh et al., 2021).

### 1.1 Experimental evidence of COVID-19 airborne transmission

A large body of evidence was reported during the course of the pandemic on different aspects of SARS-CoV-2 airborne transmission. First, SARS-CoV-2 was reported to remain viable in the air for up to 3 h, with a half-life of 1.1 h, in laboratory experiments (van Doremalen et al., 2020). In other field studies, viable SARS-CoV-2 was detected in air samples collected in hospital wards and rooms occupied by COVID-19 patients in the absence of aerosol-generating procedures (Lednicky et al., 2020; Liu et al., 2020; Nissen et al., 2020; Santarpia et al., 2020; Stern et al., 2021). The hospital environment was the focus of these studies because of the certain presence of infected subjects; nonetheless, each of these studies used different methods. Lednicky et al. (2020) collected air samples in a hospital room in the presence of two COVID-19 patients through a gentle water vapor condensation method to address the question of whether viable SARS-CoV-2 can be found in airborne particles; the study recognized the presence of viable SARS-CoV-2 in air samples collected at different distances from the patients (2–4.8 m). Liu et al. (2020) measured the concentration of SARS-CoV-2 gene copies in airborne particles in different areas of two hospitals in Wuhan. They detected very low concentrations in well ventilated environments (e.g. isolation wards and ventilated patient rooms); however, where detected, viral RNA concentration showed peaks both in the sub-micrometer and super-micrometer particle ranges. Santarpia et al. (2020) collected air and surface samples to examine the viral shedding of 13 people with COVID-19 at the University of Nebraska Medical Center and detected viral contamination of all samples. They suggested that SARS-CoV-2 environmental contamination around COVID-19 patients is extensive and hospital procedures must take into account the risk of airborne transmission of the virus. Stern et al. (2021) collected air samples of three particle sizes in a hospital in Boston (Massachusetts, US): SARS-CoV-2 gene copies were present in 9% of the samples and in all size fractions at concentrations ranging from 5 to 51 RNA copies m$^{-3}$. Positive associations were found between the probability of a positive sample, the number of COVID-19 cases in the hospital, and the cases in Massachusetts. Nissen et al. (2020) took swab samples from individual ceiling ventilation openings and central ventilation filters in COVID-19 wards at Uppsala University Hospital, Sweden. The central ventilation HEPA exhaust filters from the ward detected evidence for the presence of the virus, showing that it can be transported over long distances and that airborne transmission of SARS-CoV-2 should be taken into consideration for preventive measures.

While all of the reported studies investigated different aspects of potential COVID-19 airborne transmission, none of them provided a direct link between emissions from an infected subject and concentrations of viable SARS-CoV-2 in air – this was still considered a missing link to prove beyond doubt the airborne transmissibility of COVID-19.

### 1.2 Aims

The aim of this study was to fill this experimental gap and provide the so-called "smoking gun". To this end we have performed an experimental analysis measuring SARS-CoV-2 RNA copies in airborne particles sampled in a control hospital room occupied by an infected subject whose viral load was also measured. Experiments were performed for two different respiratory activities – breathing and speaking.

A theoretical quantification of airborne SARS-CoV-2 RNA copy concentrations reproducing the same scenarios of the experimental analysis was also performed using a novel theoretical predictive approach which is able to predict infection risk in different indoor environments via airborne transmission (Buonanno et al., 2020b, 2020a). Finally, experimental and theoretical data were compared through a metrological compatibility analysis to provide an experimental validation of the novel predictive approach. In order to estimate the metrological compatibility, the uncertainty budget for both the experimental method and the theoretical approach was calculated.

## 2 Materials and methods

### 2.1 Experimental method to measure saliva viral load and airborne SARS-CoV-2 concentration

To provide a direct link between the emission from an infected subject and the concentration of viable SARS-CoV-2 in air we conducted an experimental analysis measuring both (i) the viral load emission of the infected subject (i.e. the saliva viral load) and (ii) the airborne SARS-CoV-2 concentration (i.e. the concentration of RNA genomic copies of SARS-CoV-2 carried by airborne respiratory particles) in the infected subject's hospital room through a validated experimental method (Robotto et al., 2021a, 2021b) during two expiratory activities (breathing and speaking).

#### 2.1.1 Measurement of the saliva viral load

The saliva viral load of the infected subject (a 73-year-old man) was measured through a molecular test of the cycle thresholds (CT) for the two determined target genes, ORF8 and RNA-dependent RNA polymerase (RdRp). The viral load, expressed as RNA concentration of the transport medium ($C_{TM}$, RNA copies mL$^{-1}$), was determined through the quantitative correlations below (eqs. 1 and 2) between the SARS-CoV-2 reverse transcription-quantitative polymerase chain reaction (RT-qPCR) CT and the $C_{TM}$. The average $C_{TM}$ for the two genes can be calculated as:

$$C_{TM-ORF8\ gene} = 7 \times 10^{12} \cdot e^{-0.692 \cdot CT} \quad \text{(RNA copies mL}^{-1}\text{)} \quad (1)$$

$$C_{TM-RdRp\ gene} = 9 \times 10^{13} \cdot e^{-0.776 \cdot CT} \quad \text{(RNA copies mL}^{-1}\text{)} \quad (2)$$

The quantification curves were determined by means of a SARS-CoV-2 RNA standard (LGC Standards). The two curves (subsequently linearized to calculate the prediction intervals) were obtained with standard samples by subsequent 10-fold dilutions.

#### 2.1.2 Measurement of the airborne SARS-CoV-2 concentration

The experimental analysis to measure the concentration of SARS-CoV-2 RNA copies in airborne particles was conducted in a hospital room (70 m$^3$) of an infected subject at the Amedeo di Savoia Hospital (Turin, Italy) for two emitting scenarios – breathing and speaking. The experiments were carried out on the basis of the following procedure: i) 20 min of background measurements (with no people in the room) followed by ii) 20 min of measurements during which the infected subject was in the room just breathing (scenario A) or speaking (scenario B). The air exchange rate (AER) of the room was not measured due to logistical constrains; however, Italian standards set an AER for hospital rooms in the range 2–6 h$^{-1}$ (Decreto del Presidente della Repubblica, 1997; Regione Lazio, 2011).

Measurements were performed applying sampling and analytical methods considered reliable and validated according to the requirements of ISO/IEC 17025 (International Organization for Standardization, 2017). Further details of the sampling method and the experimental apparatus were reported in previous publications of the research group (Robotto et al., 2021a, 2021b).

Airborne particles were sampled in the room air using glass-fiber filters (one for each experiment) and a high volume sampler located more than 1.5 m from the infected subject to avoid direct exhalations and to measure the average concentration in the room (Cortellessa et al., 2021). Amongst the different methods that can be used to sample virus-laden particles, air filtration was chosen because it is effective in both capturing submicrometric particles and collecting large air volumes at the same time. Glass-fiber filters (grade MG G, 1.5-µm pore size, 10-cm diameter, Munktell Filter AB, Falun, Sweden) were adopted rather than PTFE filters or gelatine membrane filters because their porosity allows both high flow rates and good collection efficiency. The high-volume sampler employed in the experimental analyses has a flow rate of 500 L min$^{-1}$, guaranteeing high analytical sensitivity. The length of the sampling (20 min) was chosen to guarantee a negligible effect of virus inactivation as reported in our previous paper (Robotto et al., 2021b).

Once collected, the virus-laden particles underwent a subsequent elution step with Dulbecco's Modified Eagle's Medium (DMEM) or phosphate buffered saline (PBS) to extract the virus from the fiber solid matrix (Robotto et al., 2021b). The validated operational and analytical protocol applied in the present study included the following steps:
1. air sampling for 20 min (10 m³ of air sampled)
2. after sampling, the glass-fiber filters were immersed in 10 mL DMEM and transported to the laboratory at approximately 4 °C
3. the glass-fiber transport medium was supplemented with a volume of fetal calf serum up to 40% of the final volume
4. samples were subjected to the combined shaking-vortexing elution protocol described by Robotto et al. (2021b). Eighteen elution data sets were available with an average percentage of recovery of infectious virus, also defined as elution efficiency ($\varepsilon_E$), of 12.9% and a 95.4% confidence interval ranging from 2.8% to 22.9%
5. the eluate from the glass-fiber filters was then ultracentrifuged (Optima LE-80K, Beckman Coulter Life Science) for 1 h at 150000 g where the viral suspension was concentrated up to 15 times. Two measurements of the recovery efficiency of ultracentrifugation, also referred to as concentration phase efficiency ($\varepsilon_C$), were performed and ranged from 48% to 65% (median value of 57%)
6. the supernatant was discarded and the pelleted virus was concentrated in 0.35 mL of a transport medium volume, $V_{TM}$, (autoclave-sterilized PBS), and the presence of SARS-CoV-2 genomic RNA was assessed and quantified by RT-qPCR.

During SARS-CoV-2 RT-qPCR, the transport medium samples were extracted with the MagMAX™ Viral/Pathogen Nucleic Acid Isolation Kit (ThermoFisher) protocol. 200 µL of each sample were resuspended in 265 µL of inactivating solution (binding solution), then magnetic beads and proteinase K were added. The extraction procedure was carried out automatically using King Fisher Flex instrumentation. At the end of the extraction process, the RNA extracted from the samples was resuspended in 50 µL of elution solution. The eluates obtained from the previous step were analyzed in duplicate by multiplex PCR using the SARS-CoV-2 ELITe MGB Kit (ELITechGroup). The targets were RdRp and ORF8 genes specific for SARS-CoV-2 and RNase P (RP) gene as an endogenous internal control, and the volume of sample loaded into PCR was 10 µL. The PCR analysis was performed through a QuantStudio 5 Real-Time PCR System (Applied Biosystems) and the data was processed following the instructions of the PCR kit, setting the thresholds for each individual gene and evaluating the presence of suitable PCR curves. The results were expressed with the CT values for each detected target. In cases of absence of amplification (absence of the desired target), the result was reported as CT > 40. The RT-qPCR assay limit of detection (LoD) can be conservatively defined as 3 genome copies per reaction, that is, around 75 RNA copies mL$^{-1}$ of transport medium.

The average RNA concentration of the transport medium ($C_{TM}$, copies mL$^{-1}$) for the ORF8 gene was calculated through the corresponding quantification curve (eq. 1). The curve was obtained with standard samples by subsequent 10-fold dilutions; in particular, four pairs of values for both the targets were obtained (a pair of values for each dilution stage). The authors point out that for such low concentrations of $C_{TM}$, further 10-fold dilutions resulted in concentrations lower than the instrumental limit of detection. The corresponding airborne SARS-CoV-2 concentration ($C_{exp}$, RNA copies m$^{-3}$) can be evaluated as:

$$C_{exp} = \frac{V_{TM} \cdot 1000}{V \cdot \varepsilon_E \cdot \varepsilon_C} \cdot C_{TM} \qquad \text{(RNA copies m}^{-3}\text{)} \qquad (3)$$

where $V_{TM}$ is the transport medium volume to be analyzed through RT-qPCR (mL), V is sampled air volume (L), $\varepsilon_E$ is elution efficiency from the glass-fiber filters the air was filtrated through (dimensionless), and $\varepsilon_C$ is the concentration phase efficiency by means of ultracentrifugation (dimensionless). The airborne SARS-CoV-2 concentration was evaluated considering only the ORF8 gene; thus, the $C_{TM\text{-}ORF8\ gene}$ quantification curve was adopted.

The subscript "exp" was added to the average RNA concentration symbol to clearly indicate that this represents the average concentration resulting from the experimental analysis.

### 2.1.3 Uncertainty budget of the experimental method

Although an estimate of the uncertainty is mandatory for accredited clinical laboratories (ISO 15189:2012 [International Organization for Standardization, 2012]), it is rarely evaluated and reported in the publication of measuring data. This could represent a critical limitation, especially in the health sector. To evaluate the uncertainty of the average RNA copy concentration, we have applied the "Guide to the expression of uncertainty in measurement" (Joint Committee for Guides in Metrology, 2008a) to the relationship reported in (eq. 3), assuming each quantity as independent. Thus, the combined standard uncertainty ($u_{C\text{-}exp}$) can be estimated as the square-root of the linear sum of the squared standard uncertainty components, where the $i$-th standard uncertainty component is the product of the standard uncertainty ($u_i$) and its associated sensitivity coefficient ($\partial C_{exp}/\partial i$). Due to the functional relationship amongst the parameters contributing to $C_{exp}$ (eq. 3), the combined standard uncertainty can also be expressed as the product between the average concentration ($C_{exp}$) and the square-root of the linear sum of the squared standard relative uncertainty of each component (eq. 4):

$$u_{C-exp} = \sqrt{\sum_i \left(\frac{\partial C_{exp}}{\partial i}\right)^2 \cdot (u_i)^2} = C_{exp}\sqrt{\left(\frac{u_{V_{TM}}}{V_{TM}}\right)^2 + \left(\frac{u_V}{V}\right)^2 + \left(\frac{u_{\varepsilon_E}}{\varepsilon_E}\right)^2 + \left(\frac{u_{\varepsilon_C}}{\varepsilon_C}\right)^2 + \left(\frac{u_{C_{TM}}}{C_{TM}}\right)^2}$$

(RNA copies m$^{-3}$)     (4)

where $u_{V_{TM}}$, $u_V$, $u_{\varepsilon_E}$, $u_{\varepsilon_C}$, and $u_{C_{TM}}$ are the standard uncertainties of $V_{TM}$, V, $\varepsilon_E$, $\varepsilon_C$, and $C_{TM}$, respectively. For elution efficiency ($\varepsilon_E$) and concentration phase efficiency ($\varepsilon_C$), the uncertainties were evaluated on the basis of the measurement results. In particular, elution efficiency data were normally distributed, whereas for the concentration phase efficiency a rectangular distribution (ranging between the two measurements) was adopted. Thus, the corresponding uncertainties were evaluated as the standard deviation of measurements divided by the square root of the number of measurements for $\varepsilon_E$, and as the range divided by $2\sqrt{3}$ for $\varepsilon_C$.

For air volume (V), a rectangular distribution was assumed within the range provided by the instrumental specifications (± 5%), whereas for the transport medium volume ($V_{TM}$), a rectangular distribution was adopted on the basis of the laboratory standard and assuming a relative range of ± 20% of the mean value.

Finally, as mentioned above, the RNA concentration of the transport medium ($C_{TM}$) was evaluated as a derived quantity of the CT measurements through the quantification curve for the ORF8 gene. The $C_{TM}$ uncertainty was then evaluated as a deviation with respect to the linear regression of the CT values; in particular, $u_{C_{TM}}$ was estimated as the prediction interval (standard error of the prediction for N measures) of the linearized quantification curve, through the following equation:

$$u_{C_{TM}} = \sqrt{\left(\frac{1}{N} + \frac{1}{n}\frac{(CT - \overline{CT})^2}{S_{xx}}\right)\frac{SS_R}{n-2}}$$

(RNA copies mL$^{-1}$)     (5)

where:
- CT is the current CT value, obtained by means of RT-qPCR, determined as the average value of N readings (with N ranging from 1 to 3 in the present analysis as a function of the concentration, and considering that low concentrations did not allow adopting higher N)
- $\overline{CT}$ is the mean value of the CTs obtained by subsequent 10-fold dilution of the SARS-CoV-2 RNA standard
- $n$ is the number of measures (pairs) performed to obtain the quantification curve (four couples of measurements were performed in the present study due to the low concentration under investigation)
- $SS_R$ is the sum of squares of residuals
- $S_{xx} = \sum_{i=1}^{n}(CT_i - \overline{CT})^2$, where $CT_i$ is the CT corresponding to each RNA standard dilution for which the quantification curve is obtained.

The sensitivity coefficients of each input parameter ($\partial C_{exp}/\partial i$) were numerically evaluated, as stated in Annex B of the "Guide to the expression of uncertainty in measurement" (Joint Committee for Guides in Metrology, 2008b), by holding all input quantities but one (the $i$-th) fixed at their best estimates. Once the standard uncertainty $u_{C\text{-}exp}$ was estimated, the expanded uncertainty, with a 95.4% confidence interval (coverage factor of 2), was calculated as:

$$U_{C-exp} = 2 \cdot u_{C-exp} \qquad \text{(RNA copies m}^{-3}\text{)} \qquad (6)$$

## 2.2 Theoretical approach to estimate the airborne SARS-CoV-2 concentration

Recently, Buonanno et al. (2020b) proposed a predictive emission approach to estimate the viral load emission rate ($E_{vl}$) of an infected subject on the basis of the saliva viral load ($C_{TM}$, RNA copies mL$^{-1}$), the airborne particle volume concentration expelled by the infectious person during different activities (i.e. breathing, speaking, singing, etc.) ($V_d$, mL m$^{-3}$), and the flow rate expired as a function of the activity level (inhalation rate, IR, m$^3$ h$^{-1}$).

$$E_{vl} = C_{TM} \cdot IR \cdot V_d \qquad \text{(RNA copies h}^{-1}\text{)} \qquad (7)$$

The forward emission approach allowed for the first time the accurate simulation and prediction of infection risk in different indoor environments via airborne transmission both in close proximity situations (e.g. through complex computational fluid dynamics analyses [Cortellessa et al., 2021]) and in indoor environments (e.g. adopting simplified zero-dimensional models [Buonanno et al., 2020a; Mikszewski et al., 2021; Stabile et al., 2021]). Indeed, applying a mass balance approach (i.e. a zero-dimensional model which considers fully mixing conditions), the indoor SARS-CoV-2 concentration over time [$C_{theor}(t)$, RNA copies m$^{-3}$] can be estimated through the theoretical approach as:

$$\frac{dC_{theor}(t)}{dt} = \frac{E_{vl}}{V_{room}} - IVRR \cdot C_{theor}(t) \qquad \text{(RNA copies m}^{-3}\text{)} \qquad (8)$$

where the subscript "theor" was adopted to clearly indicate that this represents the concentration resulting from the theoretical approach and to differentiate it from experimentally measured concentrations. In the case of initial concentration equal to 0, the indoor SARS-CoV-2 concentration can be estimated as:

$$C_{theor}(t) = \frac{E_{vl}}{V_{room} \cdot IVRR}\left(1 - e^{-IVRR \cdot t}\right) \qquad \text{(RNA copies m}^{-3}\text{)} \qquad (9)$$

where *IVRR* (h$^{-1}$) represents the infectious virus removal rate in the space investigated and $V_{room}$ (m$^3$) is the volume of the indoor environment considered. The infectious virus removal rate is the sum of three contributions (Yang and Marr, 2011): the AER (h$^{-1}$) via ventilation, the particle deposition on surfaces ($k$, h$^{-1}$) and the viral inactivation ($\lambda$, h$^{-1}$).
Thus, the 20-min average RNA concentration ($C_{theor}$) to be compared with the measured concentrations in the two abovementioned scenarios is determined as:

$$C_{theor} = \int_T C_{theor}(t)dt = \frac{E_{vl}}{V_{room} \cdot IVRR}\left[1 - \frac{1}{IVRR \cdot T} \cdot \left(1 - e^{-IVRR \cdot T}\right)\right] \qquad \text{(RNA copies m}^{-3}\text{)} \qquad (10)$$

with T total duration of the event (20 min).
A Monte Carlo simulation (Hammersley and Handscomb, 1964) was run to estimate the RNA copy concentration in both the scenarios given a range of input values through the predictive estimation approach (Buonanno et al., 2020b). To this end, probability distribution functions and related values of the different parameters were adopted on the basis of the measurements carried out in the present study or obtained from the scientific literature as summarized in Table 1. All the parameters were assumed to be uncorrelated and Monte Carlo simulations were run performing $1 \times 10^6$ trials to estimate the average RNA copy concentration ($C_{theor}$).
For the saliva viral load ($C_{TM}$), a rectangular (i.e. uniform) distribution of the data, ranging from $2.4 \times 10^6$ to $5.5 \times 10^6$ RNA copies mL$^{-1}$ was considered on the basis of the values obtained from the SARS-CoV-2 molecular test of the subject (reported in section 3.1). For the inhalation rate, the data reported in the scientific literature (Adams, 1993; International Commission on Radiological Protection, 1994) for sitting activity levels were collected: the values ranged between 0.47 and 0.57 m$^3$ h$^{-1}$ and a rectangular probability distribution was considered.

Experimental data from the scientific literature are not definitive for the particle volume emitted ($V_d$), because the sampling method itself can affect the results due to the rapid dehydration of the large particles emitted (Abbas and Pittet, 2020; Morawska and Buonanno, 2021; Stadnytskyi et al., 2020; Yang and Marr, 2011). Still today the understanding of the initial instant of respiratory particle emission is not definitive and does not lead to conclusive answers, mainly due to the complexity of physical processes such as evaporation and the difficulty of measuring particle emissions in situ (Bake et al., 2019, 2017). As a consequence, $V_d$ was found to range from roughly $1.2 \times 10^{-3}$ to $1.2 \times 10^{-2}$ mL m$^{-3}$ for speaking ($V_{d\text{-speaking}}$, (Duguid, 1945; Evans, 2020; Johnson et al., 2011)) and from $1.0 \times 10^{-6}$ to $2 \times 10^{-3}$ for breathing ($V_{d\text{-breathing}}$, (Buonanno et al., 2020a; Morawska et al., 2009)). Rectangular probability distributions were applied to those data in the Monte Carlo simulations. The room volume (V) was measured as 70 m$^3$: volume values with a normal probability distribution were generated in the Monte Carlo simulations adopting an expanded uncertainty of 2% as estimated by d'Ambrosio Alfano et al. (2012).

AER data were generated adopting a rectangular probability distribution function with values ranging from 2.0 to 6.0 h$^{-1}$ as mentioned above. Similarly, the inactivation rate ($\lambda$) was considered equiprobable and ranging from 0 h$^{-1}$ (Fears et al., 2020) to 0.63 h$^{-1}$ (van Doremalen et al., 2020). In contrast, the deposition rate (k) is a function of the particle size; thus, on the basis of the settling velocity provided in Chatoutsidou and Lazaridis (2019) for particles < 10 μm, a log-normal distribution function with an average value of $\log_{10}(-0.62)$ (i.e. 0.24 h$^{-1}$) was considered.

Finally, for the scenario with the subject speaking for 20 min (scenario B), an actual time fraction of speaking ($TF_{speaking}$) was estimated and adopted to take into account the subject's involuntary pauses during speaking. In other words, the total length (20 min) was multiplied by a $TF_{speaking}$ ranging from 0.6 to 1.0 (rectangular distribution function). During the corresponding remaining fraction of the time (0.4 to 0, respectively) the patient was considered to be just breathing and the corresponding particle volume ($V_{d\text{-breathing}}$, particle volume emitted while breathing) was applied.

### 2.2.1 Uncertainty budget of the theoretical approach

The Monte Carlo simulations and the corresponding uncertainty budgets were carried out adopting Supplement 1 to the "Guide to the expression of uncertainty in measurement" (Joint Committee for Guides in Metrology, 2008b). In particular, the expanded uncertainty range was expressed as 95.4% confidence interval (coverage factor of 2) of the simulation results obtained from the 10$^6$ trials.

The contribution of each input parameter to the overall RNA concentration uncertainty was also evaluated by adopting a Monte Carlo simulation (instead of an analytical approach), as suggested in Annex B of Supplement 1 to the "Guide to the expression of uncertainty in measurement" (Joint Committee for Guides in Metrology, 2008b). Indeed, we have evaluated the percent contribution of each $i$-th component as:

$$\left(\frac{\partial C_{theor}}{\partial i}\right)^2 \cdot (u_i)^2 \bigg/ \left[\sum_i \left(\frac{\partial C_{theor}}{\partial i}\right)^2 \cdot (u_i)^2\right] \qquad (\text{-}) \qquad (11)$$

where ($\partial C_{theor}/\partial i$) and $u_i$ represent the sensitivity coefficient and the standard uncertainty of the $i$-th component, respectively. The sensitivity coefficients were numerically evaluated by holding all input quantities but one (the $i$-th) fixed at their best estimates, whereas the uncertainty of the $i$-th component was evaluated with a 68.3% confidence interval (i.e. adopting the standard deviation in cases of normal distributions and the range divided by 2√3 in cases of rectangular distributions).

**Table 1** – Probability distribution of the parameters used to calculate average viral indoor concentration: normal distributions were reported as average values ± standard deviation, whereas rectangular distributions were reported as median value and minimum-maximum range.

| Parameter | Distribution | Distribution parameters | references |
|---|---|---|---|
| Viral load, $C_{TM}$ (RNA copies mL$^{-1}$) | Rectangular | $4.0 \times 10^6$ ($2.4 \times 10^6 – 5.5 \times 10^6$) | Measured, this study (see section 3.1) |

| Inhalation rate while standing, IR ($m^3$ $h^{-1}$) | Rectangular | 0.52 (0.47 – 0.57) | (Adams, 1993; International Commission on Radiological Protection, 1994) |
|---|---|---|---|
| Particle volume while speaking, $V_{d\text{-}speaking}$ (mL $m^{-3}$) | Rectangular | $6.6 \times 10^{-3}$ ($1.2 \times 10^{-3}$ – $1.2 \times 10^{-2}$) | (Duguid, 1945; Evans, 2020; Johnson et al., 2011) |
| Particle volume while breathing, $V_{d\text{-}breathing}$ (mL $m^{-3}$) | Rectangular | $1.0 \times 10^{-3}$ ($1.0 \times 10^{-6}$ – $2.0 \times 10^{-3}$) | (Buonanno et al., 2020a; Morawska et al., 2009) |
| Room volume, $V_{room}$ ($m^3$) | Normal | 70.0 ± 0.7 | Measured, this study (adopting an uncertainty of 2% according to d'Ambrosio Alfano et al. (2012)) |
| Air exchange rate, AER ($h^{-1}$) | Rectangular | 4.0 (2.0 – 6.0) | Italian standard (Decreto del Presidente della Repubblica, 1997; Regione Lazio, 2011) |
| Particle deposition rate, $k$ ($h^{-1}$) | Log-normal | $\log_{10}(-0.62) \pm \log_{10}(0.30)$ | (Chatoutsidou and Lazaridis, 2019) |
| Inactivation rate, $\lambda$ ($h^{-1}$) | Rectangular | 0.32 (0 – 0.63) | (Fears et al., 2020) (van Doremalen et al., 2020) |
| Time fraction of speaking, $TF_{speaking}$ (-) | Rectangular | 0.8 (0.6-1.0) | Estimated, this study |

## 2.3 Metrological compatibility

The in-depth metrological analysis of the experimental method (section 2.1.3) and the numerical evaluation of the theoretical approach (section 2.2.1) in estimating the uncertainty of the airborne SARS-CoV-2 concentration were aimed at evaluating the metrological compatibility between experimental and theoretical data (Joint Committee for Guides in Metrology, 2008a).

The authors point out that metrological compatibility allows a conformity assessment to be made; that is, deciding if an item of interest conforms to a specified requirement. Indeed, two measurements of the same magnitude can be different, but metrologically compatible (i.e. not statistically different), if their difference falls within the experimental error considering the corresponding uncertainties or if there is a value that falls within both measurement ranges. In this paper, we apply the principles of the conformity assessment to the RNA copy concentration obtained through the experimental methodology and the predictive approach as reported in the ISO/IEC 17043 standard (International Organization for Standardization, 2010). The compatibility is assured if the absolute value of the difference of any pair of measured quantity values (mean values) from two different measurement results is smaller than the expanded measurement uncertainty (of that difference. In other words, the metrological compatibility of measurement results replaces the traditional concept of 'staying within the error', as it represents the criterion to decide whether two measurement results refer to the same measurand or not (International Organization for Standardization, 2010). Metrological compatibility is determined in terms of normalized error $E_n$ (to be less than 1), defined on the basis of the above-reported uncertainties (Joint Committee for Guides in Metrology, 2008a) as:

$$E_n = \frac{|C_{exp} - C_{theor}|}{\sqrt{U_{C_{exp}}^2 + U_{C_{theor}}^2}} < 1 \qquad (-) \qquad (12)$$

where $C$ and $U_C$ are the mean values and the expanded uncertainties referred to in the data obtained through the experimental method (exp) or the theoretical approach (theor). While the uncertainty range of the experimental data is symmetric with respect to average value, as reported in the results, the uncertainty range resulting from the theoretical approach is asymmetric with respect to the median value (see section 3.2). In view of a conservative approach, the normalized error was calculated considering the smallest value amongst (i) the difference between the highest value of the confidence interval and the median value and (ii) the difference between the lowest value of the confidence interval and the median value.

## 3 Results and discussions

### 3.1 Results of the experimental analysis and uncertainty budget

The molecular test performed on the saliva samples of the infected subject resulted in CT values for the two target genes, ORF8 and RdRp, of 21.51 and 21.41, respectively; thus the corresponding RNA concentration of the transport medium ($C_{TM}$), evaluated through the quantification curves (eqs. 1-2), ranged between $2.4 \times 10^6$ and $5.5 \times 10^6$ RNA copies mL$^{-1}$ (median value of $4.0 \times 10^6$ RNA copies mL$^{-1}$). The concentrations of airborne SARS-CoV-2 for scenarios A and B are reported in Table 2. A very low or undetectable concentration of airborne RNA copies was measured during scenario A, suggesting that the ventilation supplied in hospital rooms was effective in limiting the airborne transmission of SARS-CoV-2 when the infected person was just breathing due to the lower emissions typical of such expiratory activity. In contrast, the concentration measured during scenario B was much higher; in this case, with a speaking subject, a value of 16.1 RNA copies m$^{-3}$ was measured. This value is within the range of values (1–100 RNA copies m$^{-3}$) measured in hospital wards in previous research (Lednicky et al., 2020; Liu et al., 2020; Santarpia et al., 2020; Stern et al., 2021), confirming the presence of virus-laden particles in this type of microenvironment even beyond close proximity to an infected subject. The undetectable concentration resulting from the experiment involving the infected subject just breathing was somehow expected because the minimum detection threshold of the experimental device is around 2 RNA copies m$^{-3}$ and because during breathing, an infected subject emits about 10 times less than during speaking (Morawska et al., 2009). These data highlight the relevance of vocalization in terms of viral load emission and, consequently, of RNA copy concentration in the environment. In fact, the virus-laden particles emitted from the human respiratory tract are generated by aerosolization (i.e. by turbulent flows stripping particles from a fluid film) and by a fluid film or bubble burst process while vocal cords adduct and vibrate during speaking (Johnson et al., 2011). In particular, speaking activity was proven to emit additional particles in respect to breathing in modes near 3.5 and 5 µm, suggesting that the aerosolization of secretions lubricating the vocal chords is a major source of particles in terms of number (Morawska et al., 2009).

Table 2 – Average airborne SARS-CoV-2 concentrations ($C_{exp}$) ± expanded uncertainty ($U_{C-exp}$) resulting from the experimental analysis for both the scenarios considered. Cycle threshold (CT) and $C_{TM}$ values for the endogenous gene RP (adopted for internal control) and for the gene ORF8 (adopted to calculate the concentration through the quantification curve) are also reported.

| Experiment /scenario | Respiratory activity | CT for endogenous gene target (RP) | CT for virus gene target (ORF8) | $C_{TM}$ (RNA copies mL$^{-1}$) | $C_{exp} \pm U_{C-exp}$ (RNA copies m$^{-3}$) |
|---|---|---|---|---|---|
| A | Breathing | 38.42 | Undetermined | Undetermined | Undetermined |
| B | Speaking | 38.59 | 37.16 | 34.06 | 16.1 ± 32.8 |

The expanded uncertainty ($U_{C-exp}$, coverage factor of 2) for scenario B, resulting from the evaluation described in section 2.1.3, was equal to ± 32.8 RNA copies m$^{-3}$ (i.e. a relative expanded uncertainty of 204%). Such a huge uncertainty highlights the complexity of the RNA copy concentration measurement and underlines the importance of estimating the uncertainty value for this type of measurement: indeed, like most of the measurements in the medical field, the uncertainty of the RNA copy concentration measurement is typically not provided. The reason for such high uncertainty is expressed by the relative contributions (uncertainty weights) of the parameters contributing to the uncertainty $U_{C-exp}$. These contributions, along with sensitivity coefficients and standard uncertainties of each parameter, are summarized in Table 3. In particular, the main contributions to overall uncertainty (scenario B) are due to two parameters: $C_{TM}$, with a relative contribution of 84.9%, and $\varepsilon_C$, with a relative contribution of 13.7%. The $C_{TM}$ uncertainty is calculated by means of well-known models valid for the standard error of the prediction reported in (eq. 5); in particular, the quantification curve is based on four pairs of measures (expressed as *n* in (eq. 5)) each of which is performed with 1–3 readings (expressed as *N* in (eq. 5)). Thus the uncertainty could be hypothetically reduced only by increasing the number of standard dilutions and readings. Nonetheless this is not practically suitable when low concentrations are measured, in which case neither further 10-fold dilutions nor further readings at each dilution stage can be applied. Similarly, the mean value of the concentration efficiency $\varepsilon_C$ was obtained from just two measurements, so its uncertainty could be reduced by increasing the number of ultracentrifugation assays.

**Table 3** – Distribution of the parameters measured to calculate $C_{exp}$ and their sensitivity coefficients, standard uncertainties and contributions (weights) to the overall uncertainty ($U_{C\text{-}exp}$). Normal distributions were reported as average values ± standard deviation, whereas rectangular distributions were reported as median value and minimum-maximum range.

| Parameter | Distribution parameters | Probability distribution | Sensitivity coefficient ($\partial C_{exp}/\partial i$) | Standard uncertainty ($u_i$) | Uncertainty weight (%) |
|---|---|---|---|---|---|
| $\varepsilon_E$ | 12.9% ± 5.0% | Normal | -28.22 RNA copies m$^{-3}$/(-) | 0.0118 (-) | < 0.1% |
| $\varepsilon_C$ | 57% (48–65%) | Rectangular | -123.75 RNA copies m$^{-3}$/(-) | 0.049 (-) | 13.7% |
| V | 10 m$^3$ (9.5–10.5 m$^3$) | Rectangular | -0.0016 RNA copies m$^{-3}$/(L) | 288.68 L | 0.1% |
| $V_{TM}$ | 0.35 (0.28–0.42) mL | Rectangular | 45.96 RNA copies m$^{-3}$/(mL) | 0.04 mL | 1.3% |
| $C_{TM}$ | 34.06 (17.57–66.04) RNA copies mL$^{-1}$ | Not available | 0.472 RNA copies m$^{-3}$/(RNA copies mL$^{-1}$) | 31.98 RNA copies mL$^{-1}$ | 84.9% |

The authors point out that for the typical concentration values occurring in real life environments (i.e. < 100 RNA copies m$^{-3}$), a very large uncertainty is expected. Indeed, in that concentration range, the effect of the actual concentration on the uncertainty is negligible. This is clearly shown in Figure **1**a where the relative uncertainty as a function of the concentrations was calculated and reported: the uncertainty is extremely high (> 190%) over the entire concentration range analyzed. Analogously, in Figure 1b the RNA copy concentrations (and the corresponding uncertainty ranges) as a function of the CT values are also reported. The trend shows that, as previously reported, the measured CT value of 37.16 corresponds to a $C_{exp}$ of 16.1 RNA copies m$^{-3}$ with a 95.4% confidence range of 0–48.9 RNA copies m$^{-3}$ (i.e. $U_{C\text{-}exp}$ = ± 32.8 RNA copies m$^{-3}$) and how for lower CT values (i.e. for higher airborne concentrations) the relative confidence interval remains roughly constant.

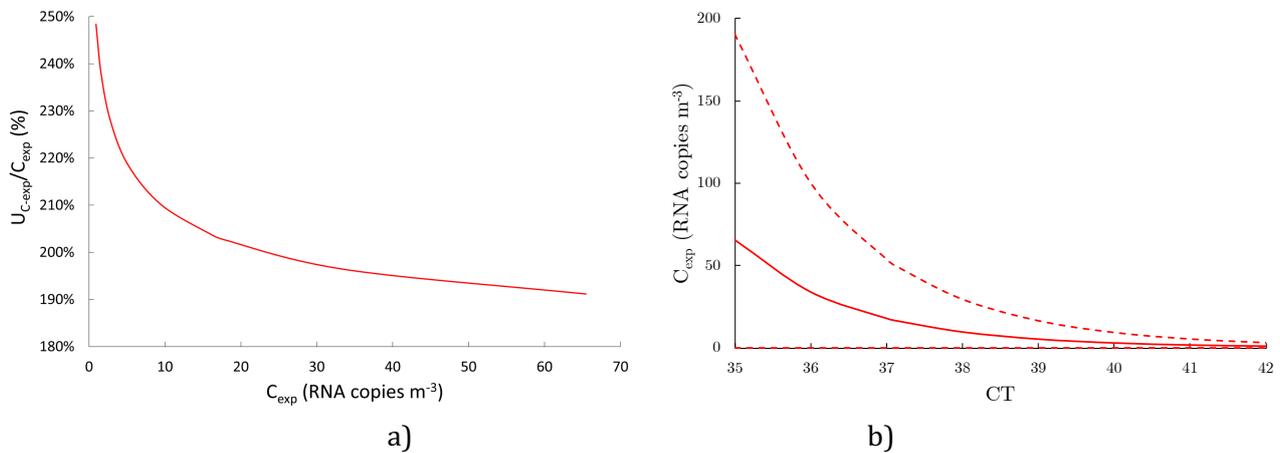

**Figure 1** – Trends of: (a) estimated relative expanded uncertainty ($U_{C\text{-}exp}/C_{exp}$) as a function of the RNA copy concentrations, and (b) RNA copy concentrations ($C_{exp}$) as a function of the cycle threshold (CT) values (solid line represents the average values, whereas dashed lines represent 95.4% confidence interval values).

### 3.2 Results of the theoretical approach and uncertainty budget

In Table 4, the average RNA concentrations ($C_{theor}$) estimated through the Monte Carlo simulations for scenarios A and B (infected subject breathing and speaking) are reported along with the expanded uncertainty range (95.4% confidence interval, coverage factor of 2). The median values were equal to 3.2 RNA copies m$^{-3}$ (with 95.4% confidence interval equal to 0.2–8.3 RNA copies m$^{-3}$) and 18.5 RNA copies m$^{-3}$ (with 95.4% confidence interval equal to 4.5–43.0 RNA copies m$^{-3}$) for breathing and speaking scenarios, respectively. The contributions of each input parameter to the overall RNA concentration uncertainty, evaluated adopting the Monte Carlo simulation as suggested by the Joint

Committee for Guides in Metrology (2008b) and described in section 2.2.1, are also reported. Such an uncertainty budget provides key information as it clearly highlights which parameter most affects the uncertainty of the RNA concentration.

The main contributions of the speaking scenario (scenario B) were the particle volume emission while speaking (75.4%) and the viral load (19.0%); the other contributions were much lower or even negligible. Additionally for the breathing scenario (scenario A), the main contributions were the particle volume while breathing (84.5%) and the viral load (13.0%). The authors point out that when the viral load is not measured (unlike the present study), its contribution to the RNA concentration uncertainty would become even larger as the $C_{TM}$ of the infected person can potentially range over several orders of magnitude (Pan et al., 2020) and, consequently, the standard uncertainty will also greatly increase.

**Table 4** – Results of indoor RNA concentrations obtained through Monte Carlo simulations for breathing (scenario A) and speaking (scenario B). Data are expressed as median values and range of the corresponding expanded uncertainties (95.4% confidence interval). The sensitivity coefficients, the standard uncertainties ($u_i$) and the contributions of the *i*-th parameter to the overall RNA concentration uncertainties are also reported.

| $C_{theor}$ (RNA copies m$^{-3}$) | | | Scenario A (breathing) | Scenario B (speaking) |
|---|---|---|---|---|
| | | | 3.2 (0.2-8.3) | 18.5 (4.5-43.0) |
| **Parameter** | **Sensitivity coefficients** ($\partial C_{theor}/\partial i$) | **Standard uncertainty** ($u_i$) | **Uncertainty weight (%)** | |
| $C_{TM}$ | Scenario A: 5.0 × 10$^{-6}$ RNA copies m$^{-3}$/(RNA copies mL$^{-1}$)<br>Scenario B: 8.6 × 10$^{-7}$ RNA copies m$^{-3}$/(RNA copies mL$^{-1}$) | 9.0 × 10$^5$ RNA copies mL$^{-1}$ | 13.0% | 19.0% |
| IR | Scenario A: 0.06 RNA copies m$^{-3}$/(m$^3$ h$^{-1}$)<br>Scenario B: 36.67 RNA copies m$^{-3}$/(m$^3$ h$^{-1}$) | 0.03 m$^3$ h$^{-1}$ | <0.1% | 1.0% |
| $V_{d-speaking}$ | Scenario A: -<br>Scenario B: 2888.1 RNA copies m$^{-3}$/(mL m$^{-3}$) | 0.0031 mL m$^{-3}$ | - | 75.4% |
| $V_{d-breathing}$ | Scenario A: 3373.6 RNA copies m$^{-3}$/(mL m$^{-3}$)<br>Scenario B: 904.5 RNA copies m$^{-3}$/(mL m$^{-3}$) | 0.0006 mL m$^{-3}$ | 84.5% | 0.3% |
| $V_{room}$ | Scenario A: -0.048 RNA copies m$^{-3}$/(m$^3$)<br>Scenario B: -0.286 RNA copies m$^{-3}$/(m$^3$) | 0.70 m$^3$ | < 0.1% | < 0.1% |
| AER | Scenario A: -0.271 RNA copies m$^{-3}$/(h$^{-1}$)<br>Scenario B: -1.235 RNA copies m$^{-3}$/(h$^{-1}$) | 1.15 h$^{-1}$ | 2.2% | 1.9 % |
| $\log_{10}k$ | Scenario A: -0.305 RNA copies m$^{-3}$/(h$^{-1}$)<br>Scenario B: -1.455 RNA copies m$^{-3}$/(h$^{-1}$) | 0.30 h$^{-1}$ | 0.2% | 0.2% |
| $\lambda$ | Scenario A: -0.305 RNA copies m$^{-3}$/(h$^{-1}$)<br>Scenario B: -1.454 RNA copies m$^{-3}$/(h$^{-1}$) | 0.18 h$^{-1}$ | 0.1% | 0.1% |
| $TF_{speaking}$ | Scenario A: -<br>Scenario B: 13.146 RNA copies m$^{-3}$/(-) | 0.12 (-) | - | 2.1% |

### 3.3 Metrological compatibility

On the basis of the average airborne SARS-CoV-2 concentrations measured through the experimental analysis ($C_{exp}$) and estimated through the theoretical approach ($C_{theor}$), and of the corresponding uncertainties ($U_{C-exp}$ and $U_{C-theor}$, respectively), the normalized error $E_n$ (eq. 12) was evaluated to assess the metrological compatibility amongst the experimental and theoretical results. In scenario B (infected subject speaking), $E_n$ was much lower than 1 (0.06), thus revealing excellent metrological compatibility. For scenario B (infected subject breathing), it was not possible to estimate the normalized error $E_n$ because the experimental concentrations were lower than the instrumental limit of detection. Thus, as mentioned above, we can infer that the experimental concentration was < 2 RNA copies m$^{-3}$; therefore, it is within the confidence interval estimated through the theoretical approach (0.2–8.3 RNA copies m$^{-3}$), thus also exhibiting metrological compatibility for this scenario.

The authors point out that metrological compatibility can be obtained (and, indeed, it was expected) when such large uncertainty ranges are estimated (both experimentally and theoretically); thus, just reporting the $E_n$ value to claim metrological compatibility could downplay the actual findings of the paper. Indeed, it should be highlighted that for scenario B (where both experimental and theoretical data are available) the difference amongst the average concentrations, i.e. the numerator of $E_n$, ($C_{exp}$ = 16.1 RNA copies m$^{-3}$, $C_{theor}$ = 18.5 RNA copies m$^{-3}$) is extremely low and it would guarantee a metrological compatibility ($E_n < 1$) even for very low, and technically unfeasible, uncertainties of ± 10%. Thus, having pointed this out, the present paper does represent an experimental validation of the predictive estimation approach for the RNA copy emission rate previously developed by the authors. Such validation is extremely important because the viral emission approach represents a major step towards predicting infection risk in different indoor environments via airborne transmission. Indeed, previous studies were performed adopting emission rates obtained from rough estimates based on retrospective assessments of infectious outbreaks only at the end of an epidemic (Rudnick and Milton, 2003; Wagner et al., 2009). In fact, such an emission approach, adopted in retrospective assessments, was able to reproduce the attack rate of a documented outbreak due to airborne transmission (Buonanno et al., 2020a; Miller et al., 2020). Nonetheless, a proper validation of the approach has not yet been performed.

The metrological compatibility assessment also highlighted the importance of estimating the uncertainty value for this type of measurement and the need to reduce it; indeed, even if a large uncertainty can be expected and accepted for simplified theoretical approaches, it cannot be tolerated for experimental measurement in the medical field. Therefore, this first-time estimation of the uncertainty (and the relative contributions) of this experimental analysis represents a key step in the process of improving the measurement method for the airborne concentration of respiratory viruses.

## 4    Conclusions

Although the detection of SARS-CoV-2 RNA copies in airborne particles has been previously reported, including in hospital wards, no studies in the scientific literature have provided a direct link between the emission of RNA copies from an infected subject and the concentration of viable SARS-CoV-2 in the air. To fill this gap, an experimental analysis was conducted under controlled conditions in a hospital room to measure both the saliva viral load of an infected subject and the SARS-CoV-2 RNA copies in airborne particles while the subject was breathing and speaking. Additionally, a novel predictive theoretical approach recently developed by the authors was also applied to the same scenarios to validate it through a metrological compatibility analysis against the experimental results. To assess the metrological compatibility for both the experimental method and the theoretical approach, an uncertainty budget was developed.

The key findings can be summarized as follows:
1. For a measured saliva viral load of the infected subject in the range of 2.4 ×10$^6$ to 5.5 × 10$^6$ RNA copies mL$^{-1}$, the corresponding average airborne SARS-CoV-2 concentration while the subject was speaking was equal to 16.1 RNA copies m$^{-3}$, whereas for breathing, the concentration was lower than the detection limit of the instrumental apparatus (i.e. < 2 RNA copies m$^{-3}$).
2. The airborne SARS-CoV-2 concentrations estimated by the novel predictive estimation approach were 18.5 and 3.2 RNA copies m$^{-3}$ for speaking and breathing, respectively, and were thus in excellent agreement with the measured values as verified by the metrological compatibility analysis.

Consequently, the following conclusions can be drawn from the experimental and theoretical results:
1. *A direct link between emission and airborne concentration was demonstrated when the subject was speaking*. The study established that the virus is airborne with consequent risk of contagion when a susceptible subject inhales infectious quanta of the virus contained in particles from respiratory activities.
2. *Inability to detect the presence of the virus in the air (in terms of RNA copies) when the subject was breathing does not mean that that there is no infection risk when people are not vocalizing*. The concentration of the virus below the method detection limit does not exclude the possibility of inhalation of infectious quanta during longer exposures in poorly ventilated environments.

3. *The link between emissions and airborne concentrations can be quantified by means of a recently developed theoretical approach.* The study validated the theoretical approach through a metrological compatibility analysis, suggesting that it can support public health authorities in deciding on control measures to lower the infection risk. For example, simulation of exposure scenarios (both in close proximity and indoor environments) enables the maximum occupancy of indoor environments under consideration (e.g., classrooms, transport microenvironments) and the maximum duration of the occupancy (events) to be determined.
4. *The complex nature of the experimental method required to measure the airborne SARS-CoV-2 concentration is unavoidably associated with high uncertainties.* The uncertainty evaluation of the experimental method conducted in this study is of great significance because, for the first time, it was possible to identify which parameters are key contributors to the uncertainty.
5. *The uncertainty budget of the theoretical approach identified the volume particle emission (if the saliva viral load is measured) as the main contributor to the uncertainty.* This means that improving the accuracy of the measurement of the respiratory particles emitted during different activities, which represents a key challenge for the scientific community involved in this type of research, could reduce the overall uncertainty of the predictive approach.